\documentclass[12pt,twoside]{article}
\usepackage{a4wide}
\usepackage[english]{babel}
\usepackage{latexsym}
\usepackage{amsmath}
\usepackage{amsfonts,amssymb}
\usepackage{graphicx}
\usepackage[round,sort&compress]{natbib}
\usepackage{fancyhdr}

\newcommand{\si}{\ensuremath{\sigma}}
\newcommand{\Zd}{\mathbb Z^d}
\newcommand{\Z}{\mathbb Z}
\newcommand{\bee}{\ensuremath{\mathcal{B}}}

\begin{document}
\begin{titlepage}
\title{Gibbsian and non-Gibbsian states at Eurandom}
\author{%
\Large{Aernout C.D. van Enter}%
  \thanks{Aernout van Enter is a member of the Steering
       Committee for EURANDOM's Random Spatial Structures programme.}\\[5mm]
  {\em Department of Mathematics, University of Groningen,}\\
  {\em PO Box 407,  9747 AG Groningen, The Netherlands} \\[5mm]
\Large{Frank Redig}
   \thanks{Frank Redig was a Senior Researcher with EURANDOM in 2001-2005, and             advisor of the RSS programme in 2005-2006.}\\[5mm]
   {\em Mathematics Institute, University of Leiden,}\\
   {\em Snellius, Niels Bohrweg 1, 2333 CA Leiden, The Netherlands}\\[5mm]
\Large{Evgeny Verbitskiy}
\thanks{Evgeny Verbitskiy was a Postdoctoral Researcher with EURANDOM in 2000-2002.}\\[5mm]
{\em Philips Research, High Tech Campus 36 (M/S 2)}\\
{\em 5656 AE Eindhoven,
The Netherlands}\\
{ and}\\
{\em Department of Mathematics, University of Groningen,}\\
  {\em PO Box 407,  9747 AG Groningen, The Netherlands} \\[5mm]
}
\date{}
\end{titlepage}
\maketitle
\begin{abstract}
We review some of the work on non-Gibbsian states of the last ten years,
emphasizing the developments in which Eurandom played a role.
\end{abstract}


\section{Introduction}
Thirty years ago some unexpected
mathematical difficulties in rigorously implementing many of the generally
used real-space Renormalization Group transformations as maps on a
space of Hamiltonians were discovered in \cite{gripea78,gripea79}.
In \cite{vEFS_JSP} these difficulties were explained by observing
that under a renormalization group map, Gibbs measures were mapped on
non-Gibbsian measures, making a map at the level of Hamiltonians
(interactions, coupling constants) ill defined.

This point of view led to further
papers covering the area of non-Gibbsianness and
Renorm\-alization--Group peculiarities
\citep{vEFS94,vEFS_JSP,vEbud,lordis95,veldis95,vEF99,fer,fer1,flr,mprf,klr,vEfhr,k1,k2,halken95,ken, leny}

Afterwards, other occasions where non-Gibbsian measures appear were found.
In particular interacting particle systems (such as stochastic Ising models)
in the transient regime were found to display a mathematically
very similar behavior \citep{vEfhr}. The main observation here is that
time evolution provides a continuum of stochastic maps, which are susceptible
to a similar analysis. In fact, if the evolution is an infinite-temperature
one (independent spin flips), then one can view it as a family of single-site
stochastic renormalization maps; and these had been already considered,
however, with a different interpretation, by Griffiths and Pearce.
Interactions can be controlled by cluster expansion techniques,
cf. also \cite{maenet}, when either time or interaction strength is small
enough.

This was probably the first contribution on the subject which
started at Eurandom.
Further developments were based on Dobrushin's programme: can one consider
non-Gibbsian measures as Gibbs measures in some more generalized sense, and
how much of the Gibbsian structure, e.g. the variational principle,
survives under such more general notions?. Moreover, 
the occurrence of non-Gibbsianness in disordered systems has been addressed, 
as well as various other examples.

In 2003, the first meeting exclusively devoted to non-Gibbsian issues was organized,
at Eurandom. One of the highlights of this meeting was the participation of Robert Israel,
who probably was the first person to clearly identify non-Gibbsianness
as the source of the Griffiths-Pearce problems \citep{isr79}. In his contribution
to the proceedings, he gave the proof (which he had announced earlier,
although it was not published before)
that in the topological sense Gibbs measures are exceptional, and that
thus non-Gibbsianness is a generic property \citep{isr92}.

The paper is organized as follows. We start with a brief description of
Gibbs states. In following sections we review  the recent work 
on preservation and loss of Gibbsianity under stochastic dynamics
and variational characterization of generalized Gibbs states. 
We end the paper with a list of open problems.

\section{Gibbs measures and quasilocality}

In this section we will describe some definitions and facts we will need about
the theory of Gibbs measures. For a more extensive treatment we refer to
\cite{geo88} or \cite{vEFS_JSP}.

We will consider spin systems on a lattice $\Z^d$, where in most cases we will
take a single-spin space $\Omega_{0}$ which is finite. The configuration space
of the whole system is $\Omega =\Omega_{0}^{\Z^d}$. Configurations will be
denoted by small Greek letters such as $\sigma$ or $\omega$, and
their coordinates at lattice site $i$ are
denoted by $\sigma_i$ or $\omega_i$. A regular (absolutely summable)
interaction $\Phi$ is a
collection of functions $\Phi(\Lambda,\cdot)$ on $\Omega_{0}^\Lambda$,
indexed by finite sets $\Lambda\subset
\Z^d$, which is
translation invariant and satisfies:
$$
\sum_{0 \in \Lambda} \ ||\Phi(\Lambda,\cdot)||_{\infty} < \infty,
$$
where $||\Phi(\Lambda,\cdot)||_{\infty}=\sup_{\sigma}|\Phi(\Lambda,\sigma)|$.
Formally Hamiltonians are given by
$$
H^{\Phi} = \sum_{\Lambda\subset \Z^d} \ \Phi(\Lambda,\cdot)
$$
Under the above regularity condition these type of expressions make
mathematical sense if the
sum is taken over all subsets having non-empty intersections with a finite
volume $\Lambda$. For regular interactions one can define Gibbs measures as
probability measures on $\Omega$ having conditional probabilities which are
described in terms of appropriate Boltzmann-Gibbs factors:
$$
\mu(\sigma_{\Lambda}\,|\,\omega_{\Lambda^{c}})
=\frac 1{Z_\Lambda^{\omega_{\Lambda^c}}}
\exp\Bigl( - \sum_{V\cap \Lambda\ne \varnothing}
\Phi(V,\sigma_{\Lambda}\omega_{\Lambda^{c}}) \Bigr)
$$
for each volume $\Lambda$, $\mu$-almost every boundary condition
$\omega_{\Lambda^{c}}$ outside $\Lambda$ and each configuration 
$\sigma_{\Lambda}$ in $\Lambda$. The expression on the right-hand
side will be denoted by
$\gamma_\Lambda(\sigma_{\Lambda}\,|\,\omega_{\Lambda^{c}})$; the collection of
$\gamma=\{\gamma_\Lambda\}$ is the \emph{Gibbsian specification}
for the potential $\Phi$.

As long as $\Omega_{0}$ is compact, there always exists at least one
Gibbs measure for every regular interaction; the existence of more than one
Gibbs measure is one definition of the occurrence of a first-order phase
transition of some sort. Thus the map from interactions to measures is one to
at-least-one. Every Gibbs measure has the property that (for one of
its versions) its conditional probabilities are continuous functions of the
boundary condition $\omega_{\Lambda^{c}}$, in the product topology.
It is a non-trivial fact that this
continuity, which goes by the name ``quasilocality'' or ``almost
Markovianness'', in fact characterizes the Gibbs measures \citep{sul,koz}, once
one knows that all the conditional probabilities are bounded away from zero
(that is, the measure is {\em nonnull} or has the {\em finite-energy}
property). In some examples it turns out to be possible to check this
continuity (quasilocality) property quite explicitly.
If a measure is a Gibbs measure for a regular interaction, this interaction
is essentially uniquely determined. Thus the map from measures to interactions
is one to at-most-one.

A second characterization of Gibbs measures uses the variational principle
expressing that in equilibrium a system minimizes its free energy. A
probabilistic formulation of this fact naturally occurs in terms of the theory
of large deviations. The (third level) large deviation rate function is up to a
constant and a sign equal to a free energy density.
To be precise, let $\mu $ be a translation invariant Gibbs measure, and
let $\nu $ be an arbitrary translation invariant measure.
Then the relative entropy density $h( \nu |\mu )$ can be defined as the limit:
$$
h(\nu |\mu ) = \lim_{\Lambda \rightarrow \Z^d} \ {1 \over |\Lambda |}
\ H_{\Lambda}(\nu |\mu)
$$
where
$$
H_{\Lambda}(\nu |\mu)= \int \log
\left(\frac
{d\nu_{\Lambda}}
{d\mu_{\Lambda}}\right)\,\, d\nu_{\Lambda}
$$
and $\mu_\Lambda$ and $\nu_\Lambda$ are the restrictions of $\mu$ and $\nu$
to $\Omega_0^\Lambda$.
It has the property that $h( \nu |\mu )= 0$ if and only if the measure $\nu$
is a Gibbs measure for the same interaction as the base measure $\mu$.
We can use this result in applications if we know for example that a known
measure $\nu$ cannot be a Gibbs
measure for the same interaction as some measure $\mu$ we want to investigate.
For example, if $\nu$ is a
point measure, or if it is the case that $\nu$ is a product measure and $\mu$
is not, then we can conclude from the statement: $h( \nu |\mu )= 0$, that $\mu$
lacks the Gibbs property.

For another method of proving that a measure is non-Gibbsian because of having
the ``wrong'' type of (in this case too small) large deviation probabilities,
see \cite{sch87}.

\section{Results on non-Gibbsian measures}
As was mentioned before, non-equilibrium models, both in the steady state
and in the transient regime have been considered. After the
papers \cite{vEfhr} and \cite{maenet} on Glauber dynamics for discrete spins,
extensions were developed in \citet{kr, lenred, derroe, KO,
vEntRus} to more general spins and types of dynamics.

Also, joint quenched measures of disordered systems have been shown sometimes
to be non-Gibbsian \citep{vEKM,vEMSS,k1,k2, vEK}, affecting the
Morita approach to
disordered systems \citep[see][]{mor,ku}. In this last case, the peculiarity can be
so strong -- and it actually is in the 3-dimensional random-field Ising
model-- as to violate the variational principle. This means in particular
that the (weakly Gibbsian) interactions belonging to the plus state
and the minus state
are different, despite their relative entropy density being zero, see
section \ref{varprin} for further discussion.
Non-Gibbsianness here means that the quenched measure cannot be written as
an annealed measure, that is a Gibbs measure on the joint space of spins
and disorder variables for some ``grand potential'', such as Morita proposed.
The Eurandom contribution \cite{klr} 
is especially relevant here.

The non-Gibbsian character of the various measures considered comes
often as an unwelcome surprise. A description in terms of an
effective interaction
is often convenient, and
even seems essential for some applications. Thus, the fact that such a
description is
not available can be a severe drawback.

\bigskip

The fact that the constraints which act as points of discontinuity
often involve configurations which are very untypical for the measure under
consideration, suggested a
notion of {\em almost} Gibbsian or {\em weakly} Gibbsian measures. These are
measures whose
conditional probabilities are either continuous only on a set of full measure
or can be written in terms of an interaction which is summable only
on a set of full measure. Intuitively, the difference is that
in one case the
``good'' configurations can shield off {\em all} influences from infinitely
far away, and in the other case only {\em almost all} influences.
The weakly Gibbsian
approach was first suggested by Dobrushin to various people; his own version
was published only later \cite{dob95,dobsh1,dobsh2}. An early definition of
almost Gibbsianness
appeared in print in \cite{lorwin92}, see also \cite{ferpfi97,maevel95, maevel97,mamore, vEV,klr, mprf}
for further developments. Some examples of
measures which are at the worst almost
Gibbsian measures in this sense are decimated or projected Gibbs
measures in an external field, random-cluster measures on regular lattices,
and low-temperature fuzzy Potts measures.

Another source of non-Gibbsian examples, which was developed at Eurandom, is Random Walk in Random Scenery \citep{HSW}.

\section{Preservation, loss, and recovery of Gibbsianity under stochastic dynamics}

Consider a lattice spin system, initially in a Gibbs state $\mu^{\Phi}$ corresponding
to a translation invariant interaction $\Phi$. This initial state is
chosen to be the starting measure of a Markovian dynamics which
has as a reversible measure a Gibbs measure $\mu^{\Psi}$ with interaction $\Psi\not=\Phi$.
The dynamics considered in \cite{vEfhr}
is high-temperature Glauber dynamics, which is informally
described as follows: at each lattice site $x\in\Zd$
the spin $\si_x$ flips at rate
$$
c(x,\si)= \exp \left(-\frac12\left(H_\Psi (\si^x)-H_\Psi (\si)\right)\right)
$$
where $\si^x$ denotes the spin configuration $\in \{-1,1\}^{\Zd}$ obtained
by changing the sign of the spin at $x$ and leaving all other spins
unchanged, and where $H_\Psi$ denotes the (formal) Hamiltonian corresponding
to $\Psi$, i.e.,
\[
H_\Psi (\si^x)-H_\Psi (\si) =\sum_{A\ni x} \left(\Psi (A,\si^x) -\Psi(A,\si)\right)
\]
By high-temperature we mean that we choose the interaction $\Psi$ to be
small and (for technical reasons) of finite range. Small is in the sense
of the norm
\[
|| \Psi ||_\alpha= \sum_{A\ni 0 } e^{\alpha |A|} ||\Psi (A,\cdot)||_\infty
\]
for some $\alpha >0$. This implies
in particular that the reversible Gibbs measure $\mu^{\Psi}$ is unique, and
that from any initial measure $\nu$, the distribution $\nu_t$ at time $t>0$ converges exponentially fast to
$\mu^{\Psi}$.

A good and intuition-guiding example to keep in mind is when
$\Phi= \Phi_{\text{Ising}}$ is the potential
of the Ising model with magnetic field $h$, i.e.,
$$
\aligned
\Phi (\{x\}, \si) &= h\si_x,\\
\Phi( \{x,y\}, \si)&= \beta \si_x\si_y
\endaligned
$$
for $ x,y$ nearest neighbors in $\Zd$, and $\Phi (A,\si)=0$ for all
other subsets $A\subset \Zd$.

The basic question addressed in \cite{vEfhr} is:
{\bf ``is the measure at time $t>0$, $\mu^\Phi_t$,
a Gibbs measure?''} In other words, is $\mu^\Phi_t= \mu^{\Phi_t}$ for some
absolutely summable interaction $\Phi_t$. The $\bee1$-norm of $\Phi_t$
\[
||\Phi_t ||= \sum_{A\ni 0} ||\Phi_t (A,\cdot)||_\infty
\]
can then be considered as a time-dependent inverse temperature.
In the case $\Psi=0$, i.e., ``infinite-temperature''
dynamics,
the limiting Gibbs measure $\mu^{\Psi}$ is a product measure, and
the dynamics then simply
consists of spins that independently (for different lattice sites)
flip at the event-times of
a mean-one Poisson process. Intuitively speaking, this
corresponds to ``heating up'' a system which starts at a finite
temperature. The question of Gibbsianness then corresponds to
the question whether
we can still associate an intermediate time-dependent ``effective temperature''
to the non-equilibrium transient states,
and how this temperature evolves. In this language, loss of Gibbsianness then corresponds to ``loss of temperature''.

To study this basic question, one considers the distribution of
the so-called double-layer system consisting
of the starting configurations, together with the
configurations at time $t$.
This is a Gibbs measure with formal Hamiltonian
\begin{equation}\label{formham}
H_t(\si,\eta) = H_\Phi (\si) - \log p_t(\si,\eta)
\end{equation}
In particular, the term $\log p_t(\si,\eta)$ is formal, and,
except at infinite temperature, a cluster expansion
(requiring $\Psi$ to be small in a strong norm)
is
used \citep{maenet} in order to see that it has the required
structure of a sum of local terms. The Hamiltonian of the double-layer system plays a fundamental role in the detection of
essential points of discontinuity of the conditional
probabilities of the measure $\mu^\Phi_t$. Roughly speaking
if $\eta$ is such that the
``random field'' system with $\eta$ the realization of the random field,
thus having  Hamiltonian $H_t(\cdot,\eta)$, has a phase transition,
then $\eta$ is a good candidate point of discontinuity (a so-called bad configuration), while if
there is no phase transition, then $\eta$ is a point of continuity (a so-called good configuration).
Natural candidates for a bad configuration in the context of the Ising model
(as a starting measure) are configurations $\eta$ which give rise
to a ``neutral'' field in \eqref{formham}, such as the alternating
configuration, or a ``typical'' random configuration, chosen
from a symmetric product measure.

The results of \cite{vEfhr}
can then be summarized as follows.
\begin{enumerate}
\item {\bf High-temperature region: Gibbsianness.}
For $\Psi$ and $\Phi$ finite range and small, the measure
$\mu^\Phi_t$ is Gibbs for all $t\ge 0$.
\item {\bf Low-temperature unbiased region: loss of Gibbsianness.}
$\Psi$ is small, has zero single-site part, and $\Phi$ is the interaction of the Ising
model with zero magnetic field at inverse temperature $\beta$.
We can choose any Gibbs measure for $\Phi$ to be the starting
measure.
Then there exists $\beta_0$ such that for all
$\beta>\beta_0$ there exist $t_0\le t_1$ such that
for $t<t_0$, $\mu^\Phi_t$ is Gibbs and for
$t>t_1$, $\mu^\Phi_t$ is not Gibbs.
\item {\bf Low-temperature biased region:
loss and recovery of Gibbsianness.}
$\Psi$ is small, has zero single-site part, and $\Phi$ is the interaction of the Ising
model with small magnetic field $h>0$ at inverse temperature $\beta$.
Then there exists $\beta_0$ such that for all
$\beta>\beta_0$ there exist $t_0\le t_1<t_2\le t_3$ such that
for $t<t_0$, $\mu^\Phi_t$ is Gibbs, for
$t_1<t<t_2$, $\mu^\Phi_t$ is not Gibbs (loss of Gibbsianness), and for 
$t>t_3$ is Gibbs again (recovery).
\end{enumerate}
It is believed that the transitions in item 2 and 3 are sharp
(i.e., $t_0=t_1$ and $t_2=t_3$) but this has not been proved, except
in the context of mean-field models (cf.\ below).
After \cite{vEfhr} there have been several further and new developments,
of which we mention the following.
\begin{enumerate}
\item {\bf Universality of short-time conservation of Gibbsianness.}
In \cite{lenred} it is proved that for arbitrary local
dynamics (including e.g.\ Kawasaki dynamics or mixtures of
Glauber and Kawasaki) and arbitrary initial Gibbs measure corresponding
to a finite-range potential, for short times the measure
remains Gibbs. The reason is that for short times,
the system consists of a ``sea'' (in the percolation sense)
of unflipped spins and isolated islands of spins where
one or more flips happened. Technically speaking, this intuition
can be made into a proof of Gibbsianness via a combination of
cluster
expansion with the Girsanov formula.
\item {\bf Interacting diffusion processes at high temperature.} In \cite{derroe} weakly interacting diffusions
are considered, and a starting measure that is high-temperature
Gibbs. In that context, via a cluster expansion technique,
Gibbsianness at all times is proved.
\item {\bf Independent diffusions starting at high and low temperatures.}
In \cite{kr} independent diffusions are considered
starting from a particular Gaussian model which can be mapped
to a discrete spin system. Here loss without recovery of Gibbsianness
is proved.

\item {\bf $n$-vector models with interacting
spin-diffusions.} In \cite{KO} and \cite{vEntRus}
it was shown that for $n$-vector models
under a diffusive single-spin
evolution the measure remains Gibbs for  short time.
If, moreover, the starting measure is at high temperature,
then the time-evolved measure will remain Gibbs forever.
These conclusions remain true, if one adds a small interaction
to the dynamics.
For the zero-field
plane rotor model in two or more dimensions, started at low temperature,  and with infinite-temperature evolution,
it is shown in  \cite{vEntRus} that the Gibbs property is lost after some finite
time, but possibly recovered after some larger time.
In $d=3$ no recovery takes place.

\item {\bf Mean-field systems with infinite-temperature dynamics.}
In \cite{kl} the Curie-Weiss model evolved via independent
spin flips is considered. In particular it is shown there that
the transitions Gibbs-non-Gibbs are sharp, and that there
is a region of parameters where the ``bad'' configurations are typical
(of measure one)
for the time-evolved measure and a region where they are untypical
(i.e., of measure zero).
\end{enumerate}

\section{Variational principle}\label{varprin}
The second part of the so-called Dobrushin's restoration programme asks for
the extension of classical results for Gibbs states (e.g., the variational principle) to the classes of generalized Gibbsian states.
Similarly, stochastic dynamics (see preceding section) or
deterministic and random transformations of Gibbs states might produce
non-Gibbsian states. Is it possible to recover a variational
principle for the transformations of Gibbs states?

\subsection{Generalized Gibbs states.}
The classical variational principle for Gibbs measures states
that if $\mu$ is a translation invariant Gibbs measure on $\Omega_0^{\Z^d}$
for a potential $\Phi$, and $\nu$ is another translation-invariant measure
with $h(\nu|\mu)=0$, then
\begin{enumerate}
\item[(a)] {\bf Specification-dependent formulation:} $\nu$ is consistent with the Gibbs specification $\gamma^{\Phi}$;
\item[(b)] {\bf Specification-independent formulation:}
$\nu$ is a Gibbs measure for $\Phi$.
\end{enumerate}

In general, $h(\nu|\mu)=0$, then, according to \cite{follmer}, $\mu$ and $\nu$
have the same local characteristics:
\begin{equation}\label{lochar}
\nu( \sigma_0|\sigma_{\Z^d\setminus\{0\}})=
\mu( \sigma_0|\sigma_{\Z^d\setminus\{0\}})
\end{equation}
for $\nu$-almost all $\sigma$. One has to take into account, that the
left-hand side is defined $\nu$-a.s., and the right hand side is defined
$\mu$-a.s. Hence, if $\nu$ and $\mu$ are two ergodic measures (and hence
singular), the interpretation of (\ref{lochar}) without further assumptions
is problematic. For example, there are measures $\mu$ such that $h(\nu|\mu)=0$
for all $\nu$. A natural assumption is that $\nu$ is concentrated
on a set of continuity points for the conditional probabilities of $\mu$.
The notion of concentration must be made explicit.

Any measure admits infinitely many (consistent) specifications. For a Gibbs measure
there is a unique {\bf quasi-local} or {\bf continuous} specification,
hence the specification which is uniquely defined everywhere, and which is
the specification of choice. For a non-Gibbsian measure we cannot
construct a quasi-local specification, and simply must choose some specification
as a reference.
Naturally, a good specification is the one close to quasi-local specifications,
i.e., the specification with a large set of continuity points.

A measure $\mu$ is called {\bf almost Gibbs}, if there exists a specification
$\gamma$ such that $\mu$ is consistent with $\gamma$ (denoted by $\mu\in\mathcal G(\gamma)$, and the set $\Omega_\gamma$ of continuity points of $\gamma$ has $\mu$-measure 1. For an almost Gibbs measure $\mu$ this specification $\gamma$
is the natural reference specification.

In \cite{flr} it was shown that if $\mu$ is an almost Gibbs
measure for specification $\gamma$, and $\gamma$ is \emph{monotonicity
preserving}, then $h(\nu|\mu)=0$ implies that $\nu\in\mathcal G(\gamma)$.
In \cite{klr}, the strong monotonicity assumption was substituted
by the requirement that $\nu$ is concentrated on
a set of continuity points of $\gamma$: $\nu(\Omega_\gamma)=1$.

If $\gamma$
is a specification and $\mu\in \mathcal G(\gamma)$, define a set
$$
\tilde\Omega_\gamma =\left\{\omega: \mu(\omega_\Lambda|
\omega_{[-n,n]^d\setminus\Lambda} )\to \gamma_\Lambda(\omega_\Lambda|
\omega_{\Z^d\setminus\Lambda})\text{ as } n\to\infty \text{ for all
finite }\Lambda\right\}.
$$
Since $\mu$ is consistent with $\gamma$, one has $\mu(\tilde\Omega_\gamma)=1$.
Moreover, if $h(\nu|\mu)=0$ and $\nu(\tilde\Omega_\gamma)=1$, then $\nu\in\mathcal G(\gamma)$ as well \citep{vEV}.
If $\mu$ is an almost Gibbs measure for specification $\gamma$ and
$\nu(\Omega_\gamma)=1$, then $\nu(\tilde \Omega_\gamma)=1$ as well, and
hence the result of \cite{vEV} can be viewed as an extension of
the corresponding result in \cite{klr}.

Nevertheless, despite the positive results mentioned above,
a specification-dependent formulation of the variational principle
has its limitations, which were identified in \cite{klr}.
Relying on a previous work on disordered systems \cite{k1, k2},
\cite{klr} provides an example of two weakly Gibbs measures
$\mu^{+}$ and $\mu^{-}$ with natural specifications $\gamma^+$ and
$\gamma^-$, respectively, such that
$
h(\mu^+|\mu^-)=h(\mu^-|\mu^+)=0$,
but
$$
\mu^+ \not\in \mathcal G(\gamma^-),\quad \mu^-\not\in \mathcal G(\gamma^+).
$$

For a recent analysis of how far one can set up the formalism, based on specifications, see \cite{Mah}.
\subsection{Transformations of Gibbs states.}

Suppose $\mu$ is a Gibbs measure on $\Omega_0^{\Zd}$ for
potential $\Phi$. There is a number of ways the state space
$\Omega_0^{\Zd}$ and hence the measure $\mu$ can be transformed:
\begin{enumerate}
\item {\bf Decimation:} For $\ell\in\mathbb N$, let $T:\Omega_0^{\Zd}\to
\Omega_0^{\Zd}$ be defined by $(T\omega)_{\mathbf n}= \omega_{\ell \mathbf n}$
for all $\mathbf n\in\Z^d$, let $\nu=T^*\mu$ be the
image of $\mu$ under $T$.
\item {\bf Single-site projections:} Suppose $\Omega_1$ is a finite set such that $|\Omega_1|<|\Omega_0|$ and $T:\Omega_0\to\Omega_1$ is onto. Let $T:\Omega_0^{\Zd}\to\Omega_1^{\Zd}$ be
defined by $(T\omega)_{\mathbf n}= T(\omega_{ \mathbf n})$ for all $\mathbf n\in\Z^d$; again, let $\nu=T^*\mu$ be the
image of $\mu$ under $T$.
\item {\bf Random transformations or Hidden Gibbs fields:}
For each $\mathbf n\in\Z^d$, $\sigma_{\mathbf n}\in\Omega_1$ chosen
independently according to $T(\cdot\, | \omega_{\mathbf n})$.
It is assumed that $T(\sigma_{\mathbf n}|\omega_{\mathbf n})>0$
for all $\sigma_{\mathbf n}\in\Omega_1$, $\omega_{\mathbf n}\in\Omega_0$,
again $\nu=T^*\mu$.
\end{enumerate}

It is known that these transformations, as well as the stochastic
transformations introduced in the previous section can produce
non-Gibbsian states. The general results on the
non-Gibbsian nature (classification of possible pathologies) of measures $\nu=T^*\mu$ or $\nu=\mu_t$ in terms of the potential $\Phi$ of
the source measure $\mu$ and the properties of $T$ ($\Psi)$ remains sketchy.
Nevertheless, it is expected that the transformed measures will admit
variational principles in some form.

In \cite{lenred1}, it was shown that a Gibbs measure $\mu$ will remain
asymptotically decoupled under Glauber dynamics for all $t>0$. Hence, for $\nu=\mu_t$, $h(\rho|\nu)$ is well-defined
for all $\rho$. This type of results is a prerequisite for a successful
variational description.

In \cite{klr}, the authors considered transformations $T$ of type (1)-(3)
and Gibbs measures $\mu$ with specification $\gamma$
under the condition that the specification $\gamma\otimes T$ is monotonicity-preserving. In this case, the image states $\nu=T^*\mu$ are
almost Gibbs for some specification $\tilde\gamma$,
and $h(\rho|\nu)=0$ implies that $\rho\in\mathcal G(\tilde\gamma)$.

In \cite{verb}, for a transformation $T$ of type (1)-(3) and
any Gibbs state $\mu$ for potential $\Phi$, it was shown that
for the image measure $\nu=T^*\mu$ one
has $h(\rho|\nu)=0$ if and only if there exists a measure $\lambda$
such that $h(\lambda|\mu)=0$ and $\rho=T^*\lambda$.
Equality $h(\lambda|\mu)=0$ by the classical variational
principle means that $\lambda$ is Gibbs for the same potential $\Phi$,
and hence, $h(\rho|\nu)=0$ implies that $\rho$, $\nu$
are transformations of Gibbs states with the same potential.

Yet another class of transformations is formed by restrictions to a layer --
the so-called Schonmann projections: let
$\mu$ be a Gibbs measure on $\Omega_0^{\Zd}$,
and $\nu$ be a restriction of $\mu$ to
a lower dimensional hyperplane $\mathbb L$, say $\mathbb L=\Z^{d-1}
\times\{0\}\subset \Z^d$. Such measures $\nu$ are often non-Gibbsian.
Nevertheless, in some cases, for example,
if $\mu$ is a plus phase of a low-temperature two-dimensional Ising
model, the corresponding measure $\nu$ can be shown \citep{klr}
to be consistent with a monotonicity-preserving specification $\gamma$,
and hence the specification-independent variational
principle for $\nu$ is valid.

\section{Conclusions and some further open problems}

In non-equilibrium statistical mechanics there are still many open questions
about the occurrence of non-Gibbsian measures.
Whether one can ascribe an effective temperature in a non-equilibrium situation
is a topic of considerable interest, (also in the physics literature, see e.g.
\cite{OP}).
The term non-Gibbsian or
non-reversible is often used for invariant measures in systems in which there
is no detailed balance \cite{lig,eylebspo,ern95}. It is an open question
to what extent
such measures are non-Gibbsian in the sense we have described here.
It has been conjectured that such measures for which there is no detailed
balance are quite generally non-Gibbsian in systems with a
stochastic dynamics, see for example \cite{lebsch} or \citep[Appendix 1]{eylebspo};
on the other hand it has been predicted
that non-Gibbsian measures are rather exceptional \citep[Open problem
IV.7.5, p.224]{lig}, at least for non-reversible spin-flip processes under the
assumptions of rates
which are bounded away from zero. The examples we have are for the moment too
few to develop a good intuition on this point, but see \cite{taslef}.

We add the remark that sometimes a dynamical description is possible in terms of
a Gibbs measure on the space-time histories (in $d+1$ dimensions).
In such cases, looking at the steady states is considering the $d$-dimensional
projection of such Gibbs measures.

\smallskip
In another direction, a study of non-Gibbsianness in a mean-field setting has
been developing.
In this case, the characterization of Gibbs measures as having continuity
properties in the product topology breaks down. For these developments, see
\cite{k3,hk, kl}.

Several problems remain open in connection to dynamics of
Gibbsian measure.
\begin{enumerate}
\item {\bf Kawasaki dynamics.} Is the transition Gibbs-non Gibbs present if we
start from a low-temperature model and
consider infinite-temperature Kawasaki dynamics
(the so-called simple symmetric exclusion process)?
Since this dynamics conserves the density of plus spins,
we should look for an initial measure that
has two phases having the same density. A possible candidate
is the Ising antiferromagnet $\Phi (\{x,y\},\si)= -\beta\si_x\si_y$ for
$x,y$ neighbors and zero elsewhere. This model has
the two checkerboard configurations $\eta_1, \eta_2$ as ground states, which have
the same density $1/2$ of plus spins. The candidate bad configuration
would then be
a checkerboard configuration of two by two squares, which has also density
$1/2$, and is neutral with respect to
the configurations $\eta_1, \eta_2$.

For the Ising model, we do not expect Gibbs-non Gibbs transitions
in the course of the evolution (because the groundstates have different
density of plus spins), but this has also not been proved.
\item {\bf Nature versus nurture transition.}
In \cite{vEfhr} is suggested that the transition Gibbs-non Gibbs
is related to a so-called nature versus nurture transition.
This is informally described as follows. Consider the Ising
model plus phase as starting measure and condition that at time $t$
a neutral configuration (such as the alternating configuration)
is observed. The question is then whether this configuration
is produced by typical path of the dynamics starting
from an atypical configuration (nature) or
by an atypical path of the dynamics starting from
a typical configuration (nurture). The second scenario is
related to the ``badness'' of the configuration.
\item {\bf Low-temperature dynamics.} If the norm of
the potential describing the dynamics is not small,
then one cannot make sense of $-\log p_t(\si,\eta)$ in \eqref{formham}
as a sum of local terms via a cluster expansion.
Therefore, this regime is still completely open.
\end{enumerate}

\section*{Acknowledgments}

Our work on non-Gibbsian issues has been to a large extent a collaborative
effort. We thank our coauthors, Roberto Fern{\'a}ndez, Alan Sokal,
Tonny Dorlas, Frank den Hollander, Roman Koteck{\'y}, Christof K\"ulske,
Arnaud Le Ny, J{\'o}zsef L{\"o}rinczi, Christian Maes, Lies van Moffaert,
Wioletta Ruszel, Roberto Schonmann and
Senya Shlosman, for all they taught us during these collaborations.
We also very much benefited from
conversations and correspondence with many other colleagues.
We thank the NDNS+ cluster for facilitating
our collaboration.
Like many of our colleagues, we have very much benefited from the
presence and activities of Eurandom. We wish it many more
happy and fruitful years!

\bibliographystyle{plainnat}

\end{document}